\newcounter{myctr}

\documentclass{ws-acs}
\usepackage{url}
\usepackage{color}

\begin{document}

\makeatletter
\def\@biblabel#1{[#1]}
\makeatother

\markboth{D. Aparicio and D. Fraiman}{Banking Networks and Leverage Dependence}

%
\catchline{}{}{}{}{}
%

\title{Banking Networks and Leverage Dependence: Evidence from Selected Emerging Countries}

\author{Diego Aparicio}

\address{Department of Economics, Massachusetts Institute of Technology, United States.}

\author{Daniel Fraiman\footnote{Corresponding author: dfraiman@udesa.edu.ar}}

\address{Departamento de Matem\'atica y Ciencias,
Universidad de San Andr\'es, Argentina.}
\address{ CONICET, Argentina.}

\maketitle

\begin{history}
\received{(received date)}
\revised{(revised date)}
\end{history}






\begin{abstract}
We use bank-level balance sheet data from 2005 to 2010 to study interactions within the banking system of five emerging countries: Argentina, Brazil, Mexico, South Africa, and Taiwan. For each country we construct a financial network based on the leverage ratio dependence between each pair of banks, and find results that are comparable across countries. Banks present a variety of leverage ratio behaviors. This leverage diversity produces financial networks that exhibit a modular structure characterized by one large bank community, some small ones and isolated banks. There exist compact structures that have synchronized dynamics. Many groups of banks merge together creating a financial network topology that converges to a unique big cluster at a relatively low leverage dependence level.
Finally, we propose a model that includes corporate and interbank loans for studying the banking system. This model generates networks similar to the empirical ones. Moreover, we find that faster-growing banks tend to be more highly interconnected between them, and this is also observed in empirical data.
\end{abstract}


\section{Introduction}
The global financial crisis that started in 2007 has stimulated an extensive literature on numerous credit-related themes such as risk assessment, financial contagion, regulatory indebtedness or liquidity ratios, misuse of derivatives (see for example~\cite{blanchard,duffie,haldane,buiter}). As the crisis unfolded, the understanding of the credit exposure at an individual bank level, as well as at an aggregate systemic standpoint became increasingly important. A given bank was perceived safer or riskier based not only on its balance sheet figures or growth estimates, but also on the financial strength of the insurance companies who would step in in the case of a debtor's default, the financial distress of the peers with whom the bank had more interbank exposure, liquidity dry-ups which could trigger massive asset fire-sales, or even the assigned probability to a lender of last resort type of bailout from the central government~\cite{gorton,brunner,klodt}.

Insurance companies played a key function in the network: they had the commitment to enter the scene if a given asset failed to make promptly payments. However once it became clear that insurance companies would be not be able to compensate for every risky asset they had insured, it also became apparent that a financial distress event could propagate across the entire financial system \cite{fender}. Banks would have to start recognizing massive asset losses, and if a bank defaulted, other interconnected institutions could follow, potentially triggering a bankruptcy cascade \cite{batti}. People fearing for its savings could line up and withdraw their deposits causing a so-called bank run, as was for example observed in the United States, United Kingdom, Iceland, Spain, or Brazil \cite{brunner,shin}.
This context is pertinent to the present work, as nowadays much of the discussion continues to be how to define and regulate higher capital requirements, while at the same time weighting the trade-offs involved, namely higher costs, or lending shifting to the shadow banking (\cite{morris2008}).\footnote{In December 2010, and as a response to the financial crisis, the Basel Committee for Banking Supervision set out the new Basel 3 Accords, even before Basel 2 or Basel 2.5 had been fully implemented by many countries \cite{basel1,basel2,basel3}. Basel 3 introduced a series of modifications, including a refined definition of bank capital, additional adjustments to the calculation of risk-weighted assets, stronger capital ratios and additional liquidity buffers.}

This paper contributes to this academic discussion by using a network-based framework to examine the structure and inner dynamics behind the links between financial institutions in different countries. The application of network theory to the financial markets \cite{heise,bastos,nier,iori}  gave place to a new set of methodologies to study interaction in interconnected agents, institutions or financial products (e.g. portfolios, stock indices, derivatives). However its literature is fairly recent, and to our knowledge none has studied the banking system structure using actual (non-simulated) data from banks' financial statements across several countries. To the extent that actual interbank data (e.g. loans, repos, swaps) is not publicly available, we argue that relevant implications can still be derived upon the common behavior of certain key balance sheet items. As we have learned from the crisis, two banks can be exposed to each other's risk, even if there is no financial transaction between them: if a bank fails, market conditions (liquidity, stock prices, mergers or acquisitions expectations) are all likely to experience cascade consequences. In general, financial institutions, be them commercial banks, investment banks, development banks, broker-dealers, or credit unions, fall under the supervision of each country's Central Bank. These institutions, henceforth referred to as banks, must usually submit their quarterly or monthly financial statements to the corresponding Central Bank regulator.

This article takes advantage of that rich, homogeneous and regular available information, and uses balance sheet data from 2005 to 2010 from five emerging countries: Argentina, Brazil, Mexico, South Africa, and Taiwan. We then construct banking functional networks from the leverage dependence across financial institutions, and analyze the structure of this networks. We find configurations that tend to group into large clusters at relatively low correlation levels. A modular structure characterized by one large bank community, some small ones and isolated banks is also found in all countries. In addition, we propose and simulate a model of corporate and interbank loans that generates rich and diverse balance sheet growth. And when we construct simulated banking functional networks according to their leverage dependence, we find results that are consistent with the empirical networks. Despite the absence of actual interbank contracts, these findings translate into relevant policy implications in terms of contagion and concentration, as well as suggest potential avenues for future research.

The paper is divided as follows. Section II describes the data and the methodology, Section III describes the main results of the paper, Section IV simulates a model of interbank loans and compares with empirical data, and Section V concludes.

\section{Methods}
\subsection{Description of the Balance Sheet Data}
 We study the interaction between financial institutions using balance sheet data from 2005 to 2010 from five developing countries: Argentina,  Brazil, Mexico, South Africa, and Taiwan.
We retrieve total assets and total liabilities for each financial institution sampled at a monthly or quarterly basis, whichever is more frequent in its reporting.
Let  $A_k(t)$ and $L_k(t)$ be the total assets and liabilities, respectively, for bank $k$ at time $t$. For each country, the number of banks $N$, fluctuates in time, and thus  $N(t)$ is a stochastic birth and death process.   Throughout the observed period 2005-2010 the existing banks at 2005 may fall bankrupt, merge with others, or simply survive until 2010. On the other side, new banks can appear in between, say in 2006, thus modifying $N(t)$ as well. As the purpose of this work is to study the interaction between banks, we shall remove from the dataset those banks with incomplete observations. For example, if a bank stops or starts reporting data in 2007, its observations are removed from that country.

Let $\tilde{N}$ be the banks that satisfy the previous condition, i.e. the effective number of banks studied. The following table shows both the number of banks in 2005 and 2011 ($N(2005)$ and $N(2011)$), as well as the number of banks present throughout the time investigated ($\tilde{N}$). Table 1 also includes the number of banks births (deaths), $N_{birth}$ ($N_{death}$), during the period 2005-2011.
\begin{table}
\begin{center}
\begin{tabular}{|cccccc|}
\hline \multicolumn{1}{|c|}{Country} & \multicolumn{1}{c|}{$N(2005)$} &
\multicolumn{1}{c|}{$N(2011)$} & \multicolumn{1}{c|}{$N_{birth}$} & \multicolumn{1}{c|}{$N_{death}$} &\multicolumn{1}{c|}{$\tilde{N}$}  \\  \hline
\multicolumn{1}{|c|}{Argentina}&
\multicolumn{1}{c|}{81}  &
\multicolumn{1}{c|}{78} &
\multicolumn{1}{c|}{3}  &
\multicolumn{1}{c|}{6}  &
\multicolumn{1}{c|}{75}  \\
\multicolumn{1}{|c|}{Brazil }&
\multicolumn{1}{c|}{105}  &
\multicolumn{1}{c|}{126} &
\multicolumn{1}{c|}{49} &
\multicolumn{1}{c|}{28} &
\multicolumn{1}{c|}{78$^*$}  \\
\multicolumn{1}{|c|}{Mexico }&
\multicolumn{1}{c|}{33}  &
\multicolumn{1}{c|}{40} &
\multicolumn{1}{c|}{14}  &
\multicolumn{1}{c|}{7}  &
\multicolumn{1}{c|}{26}  \\
\multicolumn{1}{|c|}{Taiwan }&
\multicolumn{1}{c|}{48}  &
\multicolumn{1}{c|}{38} &
\multicolumn{1}{c|}{8} &
\multicolumn{1}{c|}{18}  &
\multicolumn{1}{c|}{30}  \\
\multicolumn{1}{|c|}{South Africa }&
\multicolumn{1}{c|}{33}  &
\multicolumn{1}{c|}{31} &
\multicolumn{1}{c|}{0} &
\multicolumn{1}{c|}{2}  &
\multicolumn{1}{c|}{30$^*$}  \\
\hline
\end{tabular}
\caption{ Financial system size of the five countries studied in the period 2005-2011. $^*$One bank from Brazil and one from South Africa were excluded because of missing data in some time periods.}
\end{center}
\end{table}
For instance in Brazil 49 banks were created and 28 disappeared between 2005 and 2011.
  Only two of these new banks have gone out of the
banking system before 2011.  Except for Brazil, the rest of the countries satisfy the following equation:
 $$\tilde{N}=N(2005)-N_{death}.$$
  We divide the countries into two groups according to their size $\tilde{N}$, one composed by Argentina and Brazil, and the other composed by Mexico, South Africa and Taiwan. The first group represents a large size banking system with $\tilde{N} \approx 75$ and the second a smaller banking system with $\tilde{N} \approx 30$. Also notice the different dynamics between Brazil (also Mexico) and Taiwan: whereas in the former country the number of banks increased over time, in the latter decreased. By contrast, Argentina and South Africa exhibit very small fluctuations in $N(t)$ for the period studied.

\subsection{Network construction}

The construction of a network of associations out of balance sheet variables requires selecting an interaction metric that meets two general properties. First, it should be robust to spurious correlations so that we avoid as much as possible to use a non-stationary variable that may increase or decrease simply following a global trend. And second, it should be relevant to describe an economic process. In other words, if two banks are associated with a given variable, it should mean that they are subject to certain equivalent market or economic conditions, and therefore may share the same risk. We argue that balance sheet aggregates such as assets are inappropriate to define a network, and instead propose to use the leverage ratio.

Consider for instance total assets, which are regarded as a proxy of bank size, and are subject to a large number of global or macroeconomic circumstances. Under normal market conditions, the assets of a financial institution are expected to grow with time, as a result of general growth of the economy, inflation, balance sheet effects from expansionary monetary policy, population (customer base) growth, increased access to banking services (banking penetration), higher value of the securities held by the bank (stocks or debt). It seems that almost any pair of banks, regardless whether they are contractually linked or not, will yield a positive assets' correlation, and thus will incorrectly suggest an overlinked network.

 \begin{figure}
\hspace*{-0.5cm}\includegraphics[width=1\textwidth]{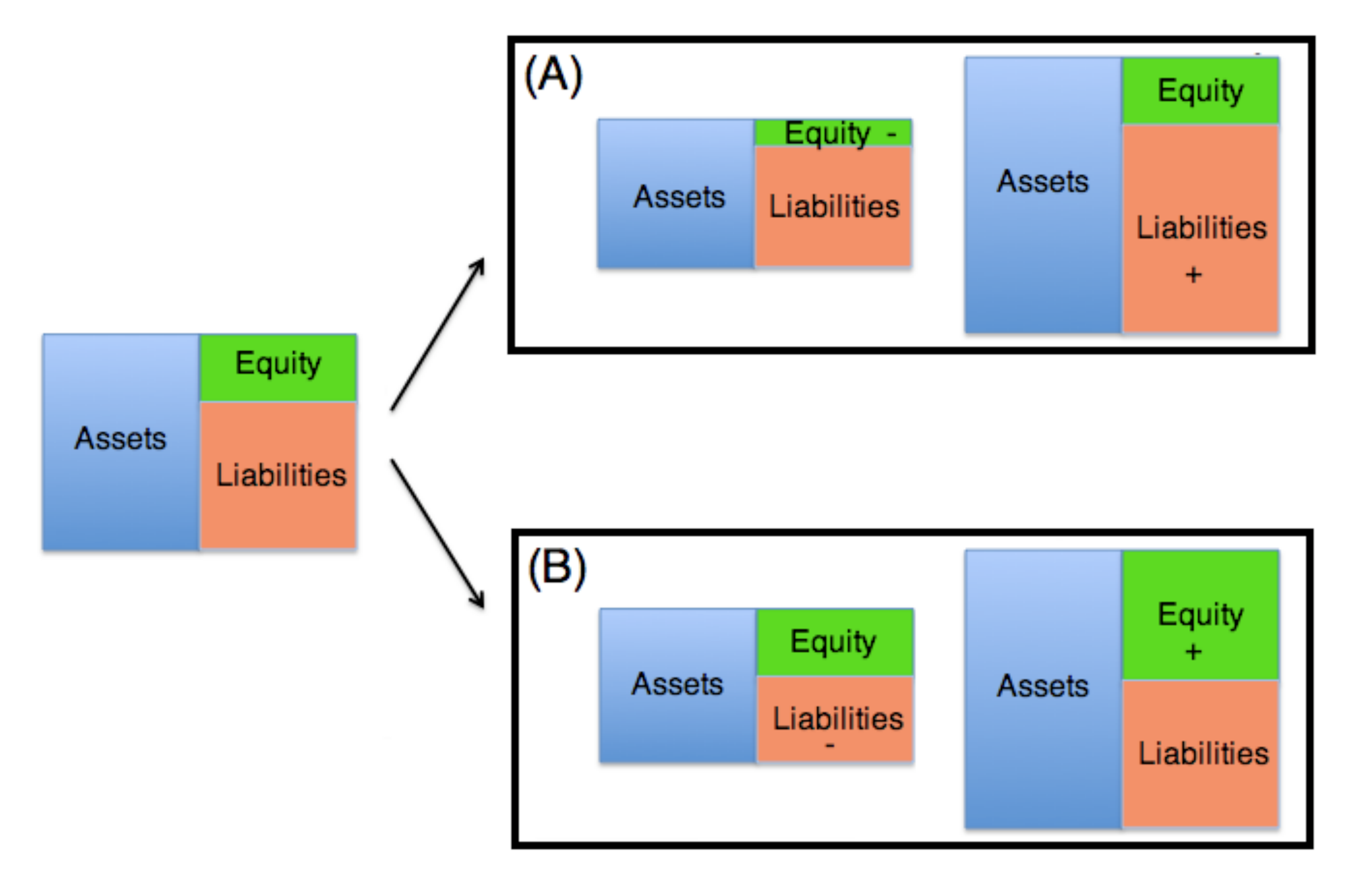} 
\caption{\textit{Two examples of the evolution of a bank's leverage}. Increase (A) and decrease (B) of the leverage ratio. }\label{fig1}
\end{figure}

We instead propose using the degree of indebtedness, henceforth referred to as the leverage ratio, as the metric on which to base the interactions. We define the leverage as follows
$$\mbox{$Leverage$} = \frac{\mbox{$Liabilities$}}{\mbox{$Equity$}}$$

Where liabilities and equity refer to the balance sheet variables of a given bank which must satisfy the usual accounting identity: $Assets=Liabilities+Equity$, and as measured at book value. Therefore, the leverage ratio of bank $k$, $Leverage_k=\frac{L_k}{ A_k-L_k}$, indicates what fraction of equity and debt this bank is using to finance its assets.\footnote{This ratio distinguishes from the capital ratio in that the capital of a bank is composed by common equity (less goodwill) as well as Tier 1/2 components such as hybrid securities, convertible capital instruments, preferred stock, provisions, loan loss reserves, and other adjustments. This definition also varies from country to country.}

Consider for instance Panel (A) and Panel (B) of Fig. 1, which shows an example where the leverage increases and decreases, respectively. The former may happen, for instance, through a credit quality shock, such that the value of the existing claims decreases and the bank increases its loan loss provision. Such loan write-offs translate into a reduction in total equity, thus increasing the ratio. The leverage can also rise if, for example, the bank simply issues debt to finance a potential acquisition or expand its credit portfolio. What regards to Panel (B), a bank may deleverage by prepaying debt with its own cash, where both total assets and liabilities decrease. It is also possible for the shareholders to conduct a capital increase, with proceeds either paying down debt or cushioning its assets. These examples illustrate only a few of the many different scenarios why a leverage ratio may fluctuate within a given financial institution.

Finally, to address the stationary concern Fig. 2 depicts the median (results remain unchanged with the average) leverage ratio for each country as a function of time. Contrary to what happens with liabilities and assets (Fig. A.1), the leverage ratio does not exhibit any strong time dependence. Each country exhibits its own leverage range, with more or less fluctuations over time. Taiwan, for example, is characterized by exhibiting a large median leverage ratio ($\sim 16$) while Argentina and Brazil exhibit a lower ratio ($\sim 6$).\footnote{As fas as we investigated, the size of the banking system is not related to the average leverage ratio.}

Leverage levels across countries are expected to differ due to different bank capital regulation. Each country may impose its own restrictions on hedging operations, ability to buy or sell foreign currency, place debt in the national or international markets (and at which currency), banks may be subject to a particular government program (e.g. subsidized mortgage or consumer credit programs with predetermined interest rates), restrictions on dividend distribution. The Central Bank may also have in place a monetary and/or inflation target rate, which in turn modifies the reference rate at which banks lend each other and hence the rates at which households or corporates access to credit. Countries also differ in their accounting consolidation practices or taxation. Overall these factors limit the construction of leverage-based networks at the country level, as opposed to aggregated inter-country networks.

In particular, we construct country-level networks of leverage dependence as follows.
\begin{enumerate}
\item[1.] Compute the leverage time series for all banks from 2005 to 2010 in a given country. After accounting for mergers and acquisitions, banks with incomplete data points are excluded (either because they appeared later or disappeared sooner than the 2005-2010 range).
\item[2.] Compute the leverage correlation matrix between all pairs of banks.
\item[3.] Establish links between banks (nodes) whenever the correlation exceeds a given threshold $\rho$. The structure of the network is studied for different values of $\rho$.
\end{enumerate}

Finally, notice that the leverage ratio also presents the important advantage of not depending on the currency, thus allowing us to use the same variable across countries. Furthermore, any difference in regulation or local practice of an improved indebtedness ratio (e.g. capital ratio after Tier 1 / 2 adjustments) is neutral within a country (for it applies to all) and not relevant across countries, since this paper does not compare absolute leverage levels,  but rather the interaction that can be derived from common leverage behaviors.

 \begin{figure}
\hspace*{-0.5cm}\includegraphics[width=1\textwidth]{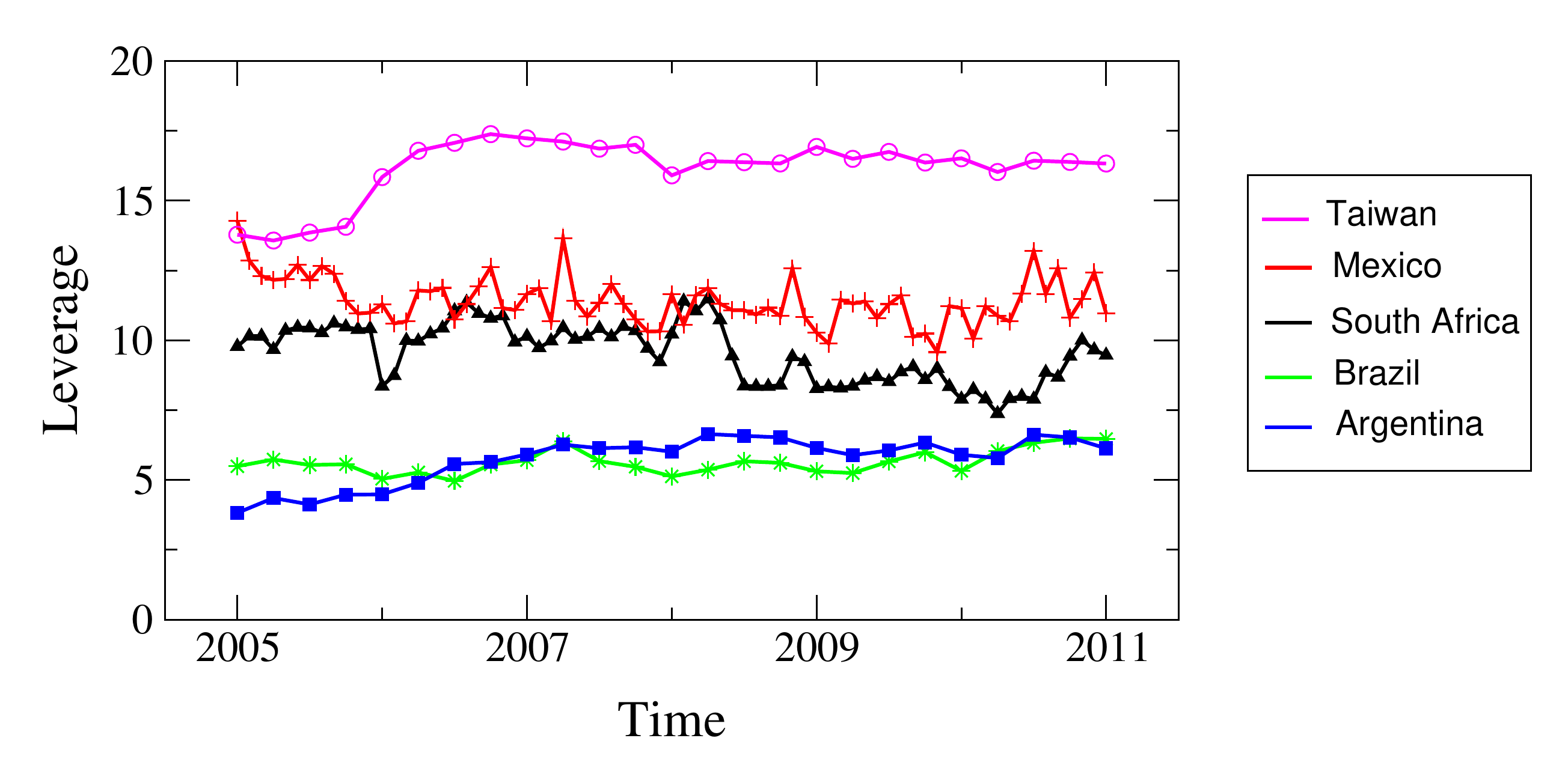} 
\caption{\textit{Stationary behavior of the leverage ratio.} Median leverage as a function of time. Each color corresponds to a country. Similar results are obtained using the mean as measure of central tendency. The overall-time median leverage is 16.4 for Taiwan, 11.3 for Mexico, 9.7 for South Africa, 6.0 for Argentina, and 5.5 for Brazil.}\label{}
\end{figure}

\section{Results}
\label{sec:results}

We next illustrate an example using data from the Argentine banking system. Given two time series, corresponding to the leverage evolution of two banks, $x(t)$ and $y(t)$ with $t=\{1,2,\dots, T\}$, we compute the Pearson correlation coefficient.
Where this coefficient takes value $r \in$ [ -1, 1] , and is equal to 1 when the two time series are identical ($x(t)=y(t)$ for all $t$).
\begin{figure}
\hspace*{-0.5cm}\includegraphics[width=1\textwidth]{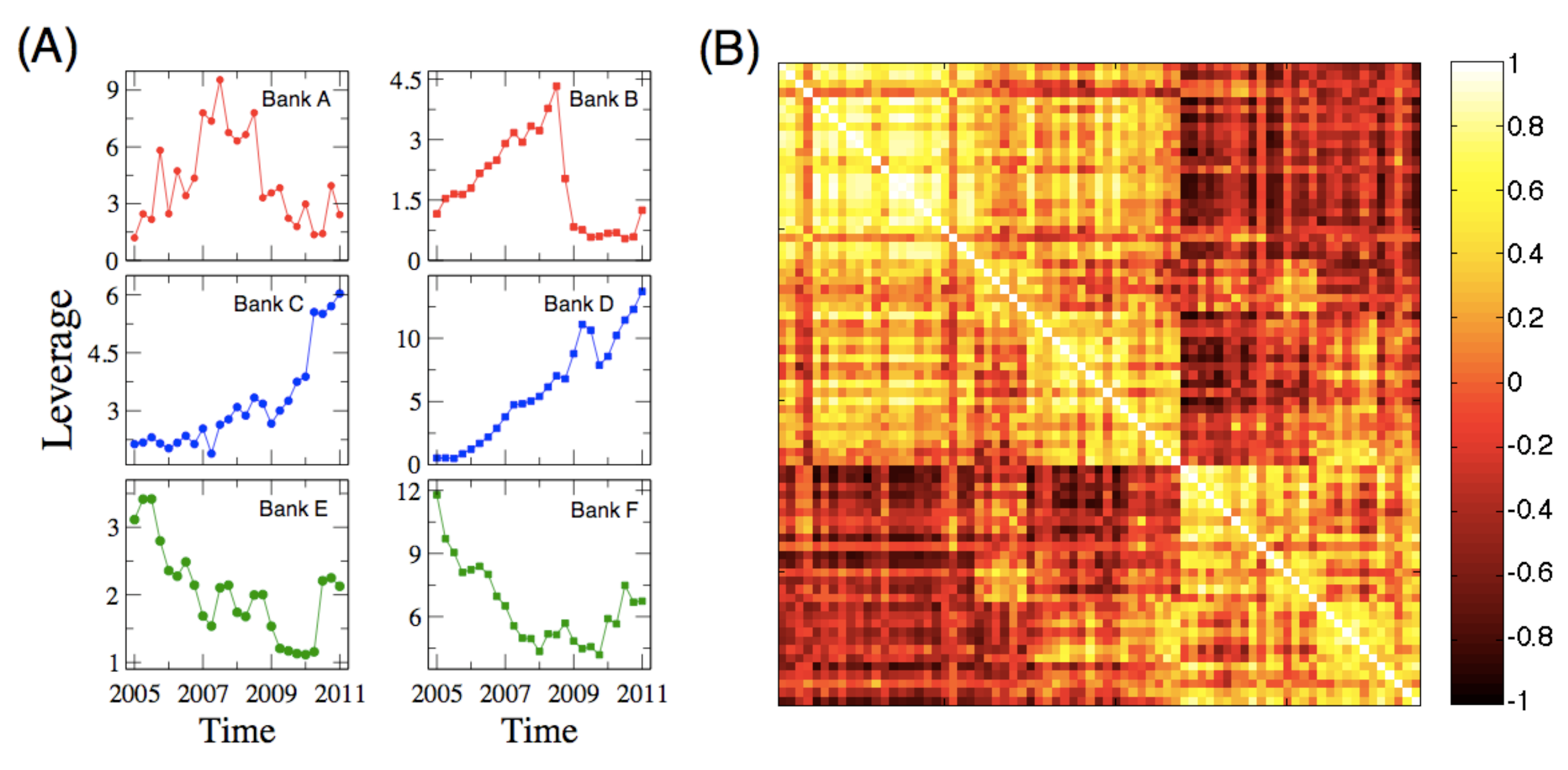} 
\caption{\textit{Banks' leverage dynamics .}  Upper, middle, and lower panels show characteristics leverage curves from different banks in Argentina.
The correlation between the two curves in each panel is shown on the right. Note that there are cases of negative correlation, e.g. between the red curves of the middle and low panel is -0.9  and between black curves is -0.86.} \label{}
\end{figure}

The correlation matrix is composed by the $\tilde{N}(\tilde{N}-1)/2$ pairs of banks correlation coefficients (interactions). Figure 3 (A) shows the leverage ratio evolution for six  (from a total of 75) different Argentine banks. Three colors are used to illustrate the different behaviors observed throughout the sample. For example the middle (bottom) two leverage curves in blue (green) correspond to two banks that exhibit an increasing (decreasing) leverage evolution. Another type of leverage evolution is shown in red, with a concave-shaped leverage behavior. The dynamics of the Argentine banking system is found to be very rich from an indebtedness perspective. As we shall see later, these rich dynamics lead to a complex banking network.

It also follows that banks C and D have a similar leverage increasing pattern. The same is found between banks E and F with practically a scale factor of difference between them, and to banks A and B, although with a different type of evolution. Each of these pairs of banks exhibit a large correlation coefficient between its corresponding leverage time series, e.g. $r= 0.81, 0.84$, and $0.83$ for the pair of banks A-B, C-D, and E-F respectively. However, if we compute the leverage correlation between bank C (or D) and E (or F) this value will be negative, thus stating that the two banks exhibit an opposite behavior. In this last case the four correlation coefficients range between -0.35 (for C-E) and -0.65 (D-E). The remaining possible interactions between the six banks fluctuate around zero.

The complete characterization of interdependence between the 75 Argentine banks is captured by the correlation matrix in Panel B, with each column and row representing a bank. A special algorithm is used to order the banks for a better visualization, such that light yellow ``spots'' correspond to group of banks that share a similar behavior among them (large $r$ coefficient), while negative correlations are represented in dark red color.

Although the leverage dependence does not imply an interbank transaction (e.g. as would be a repo agreement), a strong and recurring synchronization is still relevant in the understanding of concentration and systemic risk. It is reasonable to think that a highly leveraged network where many banks are simultaneously leveraged does not imply the same contagion risk as does a network where only a small set of banks are moderately leveraged. We can also think of the degree of negative and positive correlations as another determinant of how robust the banking system is. In this sense, identifying and studying which banks form clusters that tend to leverage (deleverage) all together can also provide useful policy implications at a time where stricter banking regulation is at the core of the debate (\cite{morris2008}).

Interestingly, from a risk-sharing point of view, it would be optimal for the financial institutions not to share identical leverage dynamics, such that a shock affecting the indebtedness of a given bank is not be followed by its peers. For a banking system to remain healthy and resilient to financial contagion, its dynamics needs to be heterogeneous. That is, greater heterogeneity helps to isolate individual shocks by setting a less interconnected financial system. Further, this can be understood as the system's endogenous solution to reduce its risk from financial shocks.

Notice that the economic implications behind the positive and negative dependence scenarios are quite different. A positive correlation between two banks implies that they both increase and decrease their debt-to-equity ratio simultaneously. In this case, we would expect both banks to share (or target) similar growing. Alternatively, banks' balance sheets could be affected by exogenous policies along the theories of balance sheet effects (\cite{kiyotaki,krugman,cespedes2004}), bank lending channels (\cite{mishkin1995,kishan2000}), and bank runs (\cite{galiani2003,calvo2003,amato2002}).

On the other hand, a negative correlation implies that when one bank increases its leverage, the opposite is true for the other bank, and vice versa. A potential explanation for this case is related to the literature of fly-to-safety, i.e. in periods of financial turmoil households withdraw their deposits from a small bank and transfer them to a bigger (perceived as safer) institution. Thus liabilities decrease in the small bank, but increase in the other bank, resulting in a negative leverage correlation between them. As documented in \cite{oliveira2011}, this effect was observed in Brazil in late 2008, where depositors ran from smaller banks to larger banks, the latter ones perceived as too-big-to-fail institutions with implicit guarantees from the Central Bank.

An alternative mechanism comes from what the literature calls the depositors discipline: depositors may punish banks with poor performance either by demanding higher interest rates or by withdrawing their deposits. \cite{schmukler2001} find evidence in Argentina, Chile, and Mexico of the depositors discipline mechanism, in particular that bank deposits growth falls as risk exposure increases. Also \cite{maechler2006,imai2006} and \cite{flannery} document that depositors favor big banks, although it is hard to differentiate whether depositors base their decisions on too-big-to-fail sentiments or bank fundamentals. Also, \cite{yeyati2004} report that during the 2001 Argentine convertibility crisis deposit withdrawals were more pronounced in banks with higher risk taking, thus providing further evidence of opposite changes in balance sheet accounts across banks.

 \begin{figure}
\hspace*{-0.5cm}
\includegraphics[angle=0,width=1\textwidth]{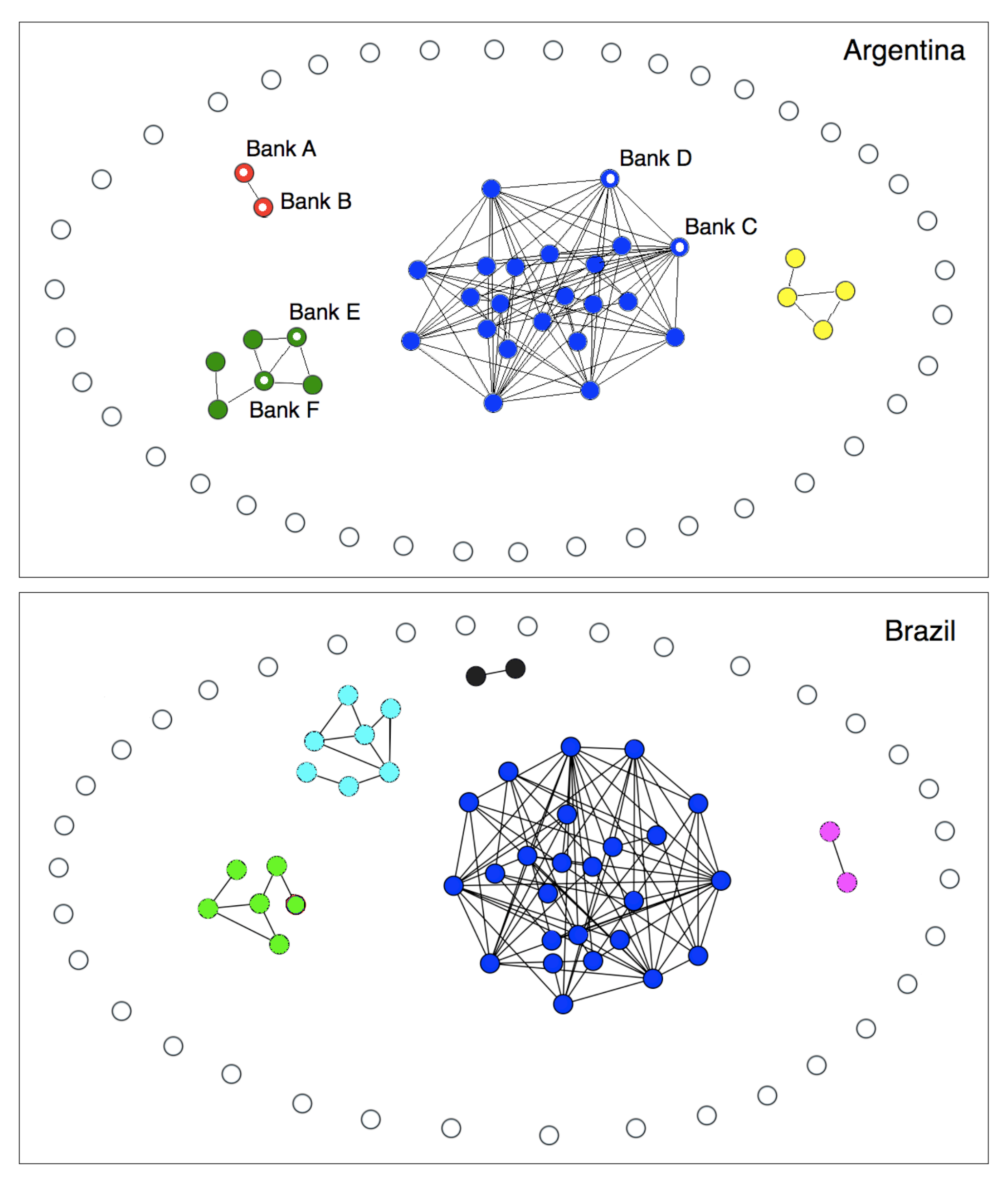} 
\caption{\textit{ Banks' leverage network for two large size banking systems.} Upper (lower) panel corresponds to Argentina (Brazil). In both countries $\langle k \rangle=2.5$.    } \label{}
 \end{figure}

We now characterize the complete Argentine network. The upper panel of Fig. 4 depicts the graphical representation of the Argentine financial network with $M$ links, or with an average degree $\langle k \rangle$ equal to $2M/\tilde{N}$. Recall that each node represents a bank, and the $M$ highest banks' pair correlations are connected by an undirected edge. Put alternatively, an edge (or link) between two nodes (or vertices) is added whenever its leverage pair correlation exceeds a certain threshold value $\rho$. The same banks showed in Fig. 3 are highlighted in the network representation of Fig. 4. Fig. 4 shows clusters of banks, where a cluster is defined as a community of interconnected (linked) banks. And each color represents a different leverage (temporal) dependence. In particular, the network can be characterized by four homogeneous groups: a large cluster of banks that essentially have an increasing leverage evolution (e.g. banks C and D); three small groups with various leverage behaviors (decreasing, concave-shaped, and a mixed of both); and finally a large set of isolated nodes (41 from 75), represented with white circles.

Interestingly, we also find that the modular structure observed in the Argentine banking system is similar to other countries. Despite having different size, regulation, or even average indebtedness level, the leverage dynamics are consistent across countries. Bottom panel of Fig. 4 shows the Brazilian banking network using the same methodology as well as the same $\langle k \rangle  = 2.5$ used in Argentina. It can be observed that Brazil' s network is similar to Argentina's, namely both are characterized by a large community of banks, another set of small clusters of banks, and many isolated banks.

As the correlation threshold decreases (and the number of links $M$ increases), both the isolated and the small-cluster nodes start merging into one large interconnected cluster. This merger-type of behavior between different groups can be observed in Fig. 5, where the fraction of nodes that belongs to the largest cluster (blue nodes as per Fig. 4) is shown as a function of the correlation threshold. As we lower the threshold $\rho$, the largest cluster starts absorbing smaller communities. Big jumps in this graph indicate that two large clusters have merged. For example, in Brazil's case a small change in the correlation threshold around 0.6 causes the two largest clusters (each one representing approximately 40\% of the nodes) to be either merged or divided.  In Argentina this similar and sudden concentration occurs at a greater correlation threshold ($\rho=0.75$). We also find that in both countries 90\% of the banks belong to a unique cluster at a relatively small correlation threshold $\rho=0.5$, thus suggesting that the financial networks may become suddenly too concentrated.

Finally, we replicate the analysis to the smaller banking systems of Mexico, South Africa and Taiwan. Fig. 6 (A) shows the leverage-based networks of these countries for an equal number of links per node,  i.e. for  $\langle k \rangle=2.5$. Similar network properties are observed across these countries. In particular, although Taiwan's banking system is more volatile (large fluctuations in $N(t)$), the banking network structure is very similar to the South African one, and both have almost the same total number of banks ($\tilde{N}=30$). This result is strengthen by the similar relationship between the size of the largest cluster and the correlation threshold shown in panel B.

The Mexican network, on the contrary, seems to exhibit certain differences with respect to the other two. For $\rho = 0.5$ the largest cluster does not have significant weight (right bottom panel), as it only accounts for 30\% of the total nodes in the network. As the correlation threshold decreases, a second cluster grows until is finally merged with the largest one (left bottom panel). Thus we observe that the consolidation of the Mexican banking network is characterized by the merger of two large clusters, whereas in the other countries there is only one large cluster which then absorbs either individual nodes or very small clusters.

 \begin{figure}
\hspace*{-0.5cm}
\includegraphics[angle=0,width=1\textwidth]{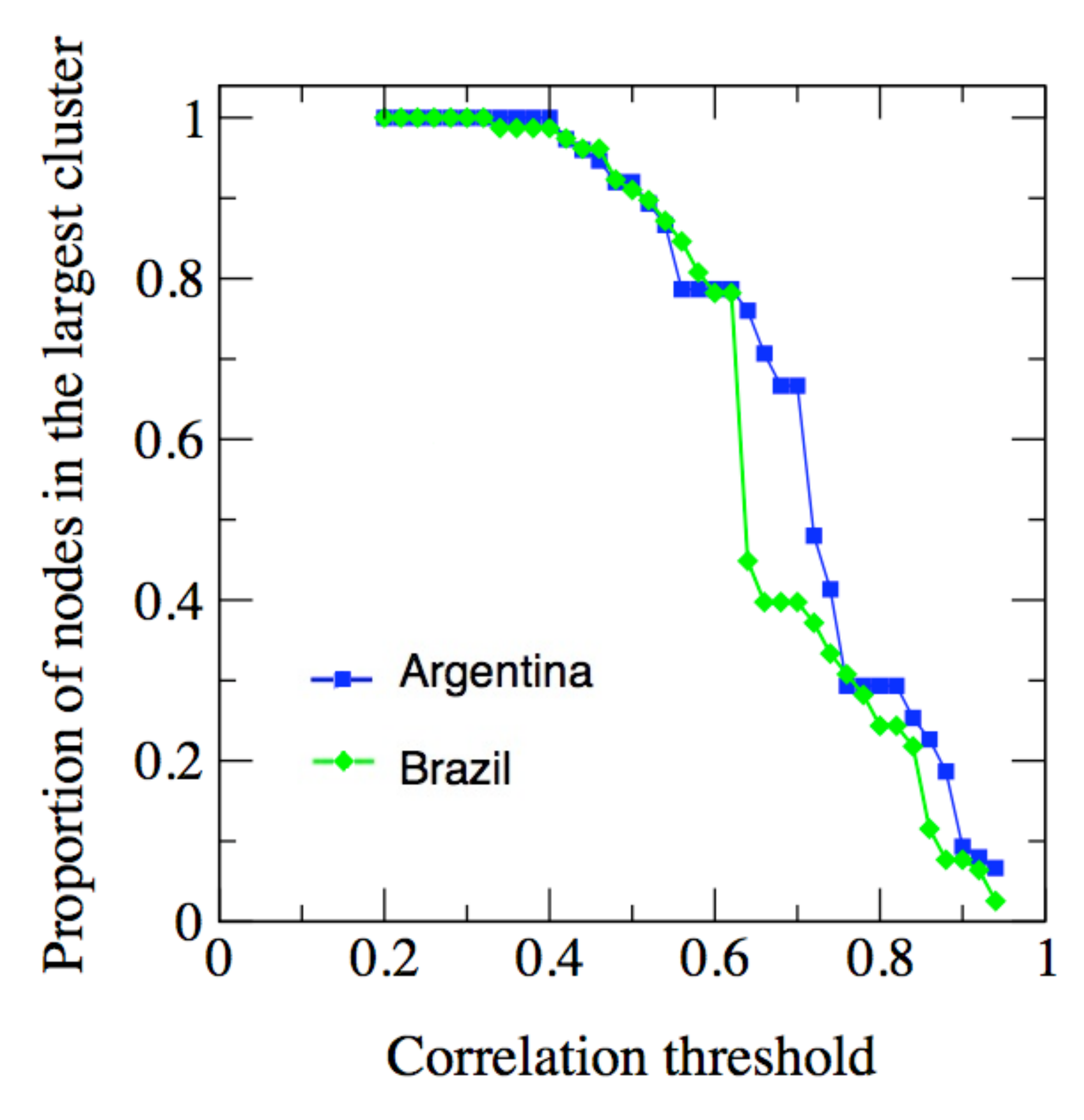} 

\caption{ \textit{ Size of the largest cluster.} Proportion of banks that belongs to the largest cluster as a function of the correlation threshold.
 } \label{}
 \end{figure}

In summary, our proposed method to examine the banking system from a leverage perspective yields networks composed by a large group of banks that are densely interconnected (blue cluster in Fig. 4 and 6), a few smaller clusters, and many isolated banks. Interestingly, this topology is homogeneous across countries, as depicted in Fig. 4 and 6. As the interaction criterion becomes more and more lax (smaller $\rho$, in absolute value), banks start re-grouping into a unique financial cluster. Furthermore, the network structure is not dominated by one hub, as it occurs with many nature networks (\cite{newman,caldarelli}). On the contrary, these banking networks exhibit a modular structure where each module is densely connected. Only a few banks function as \textit{bridges}, or connectors between different modules.

The existing structure also suggests a somewhat fragile property: a small modification to these connector banks can have sensible effects to the configuration of clusters, as well as to the isolation or propagation of a shock across the network. On the one hand, highly interconnected banks (hubs) are not desirable, since a failure in a hub can be spread out the network, potentially causing an aggregate system breakdown. But on the other hand, some degree of interconnectedness is positive, as it helps to reduce the exposure of an affected node by sharing and loading off risk with its peers. Therefore we argue that the existence of dense network structures may be desirable to some extent in order to increase risk sharing and reduce hub exposure. In this sense, in order to reduce risk and increase shock resilience, regulators should recognize a basic trade-off: as connectivity and clustering increases, individual (idiosyncratic) risk may decrease due to better risk sharing, but global or systemic risk could increase thanks to faster channels of propagation. The degree of optimal bank connectivity, and how can it be reached, are still active questions in the literature.

 \begin{figure}
\hspace*{-0.6cm}
\includegraphics[angle=0,width=1.05\textwidth]{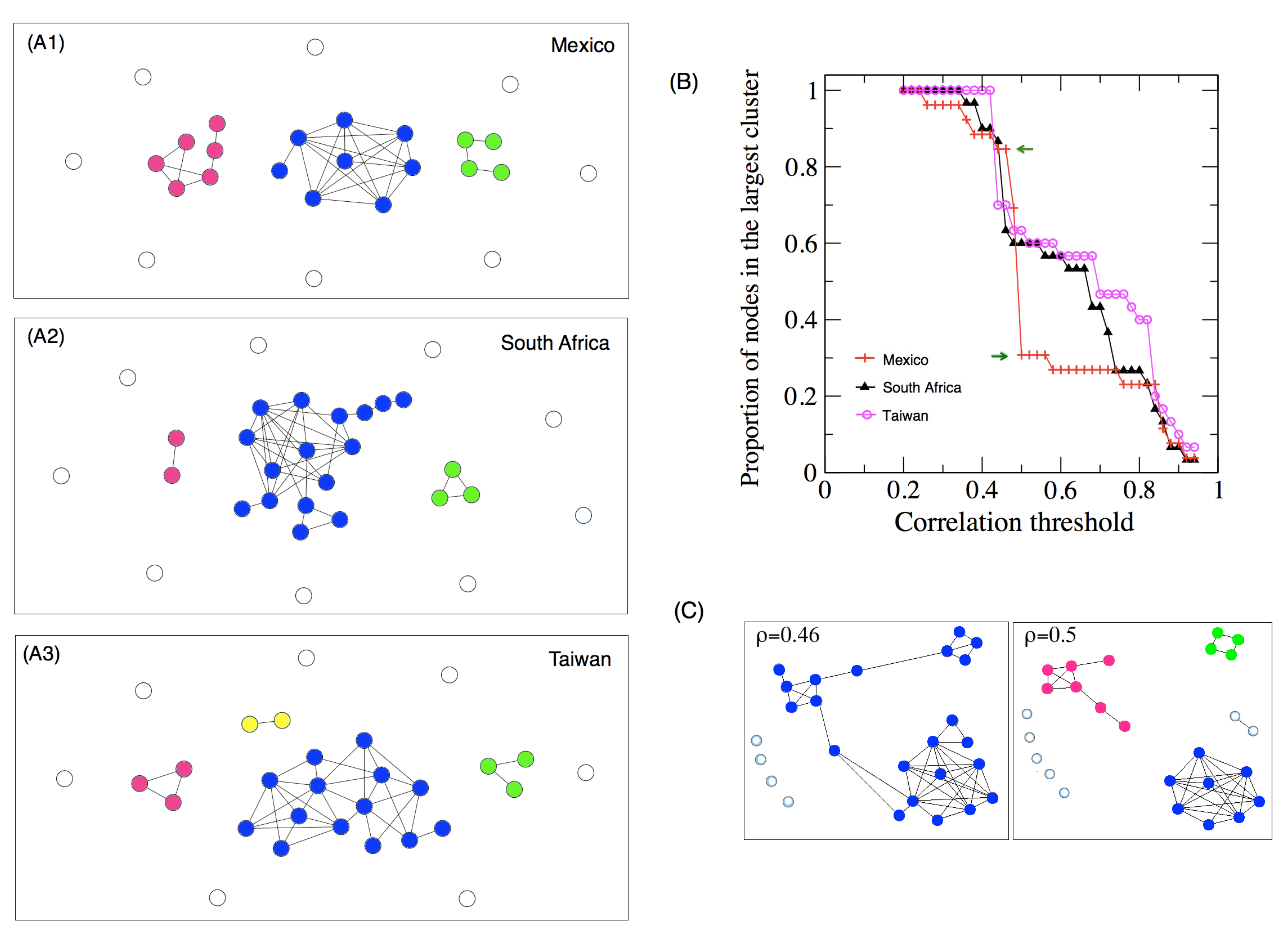} 
\caption{ \textit{Banks' leverage network for three small size banking systems, and the size of the largest cluster.} 
(A) Mexico', Taiwan', and South Africa's networks. (B) Proportion of banks that belongs to the largest cluster as a function of the correlation threshold.
(C)  Mexican network at two different correlation thresholds, as indicated by the two green arrows in the upper panel.
 } \label{}
 \end{figure}

\section{A Model of Corporate and Interbank Loans}

\subsection{Description}
In this section we intend to contrast the empirical results from Section \ref{sec:results} with simulations from a model of corporate and interbank loans. We first start by describing the set up of the model. We propose that the corporate sector requests loans from the banking system according to a given random process, and links between banks arise due to insufficient liquidity to meet these loans. This lending effect generates rich balance sheet and leverage dynamics that yield networks that are overall consistent with the empirical results.

Our model builds on the interbank model proposed in \cite{raberto}. For simplicity we assume perfect information in the economy, although as explained by \cite{chen1999} imperfect information can result in an additional source of financial risk. Further, although we think of agents in the real sector as the implicit source of interbank links, the model does not generate balance sheet dynamics from the agents' side of the economy. See \cite{lengnick} for an example of an agent-based interbank network model. See also \cite{lebaron}  for a review of ACE (agent-based computational economics) modeling in economics.

The model is structured as follows.

\begin{enumerate}

\item[1.] The banking system is composed by $N$ banks. Assets are defined as the sum of liquidity $L$, iliquid assets $I$, corporate loans $C$, and interbank loans $B^{L}$. Liabilities are defined as the sum of deposits $D$ and interbank debt $B^{D}$. Moreover, any bank $i$ at time $t$ must satisfy:
$$Assets=Liabilities+Equity$$
$$L^{i}_{t}+I^{i}_{t}+C^{i}_{t}+B^{L,i}_{t}=D^{i}_{t}+B^{D,i}_{t}+E^{i}_{t}$$

\item[2.] Let assets $A^{i}_{1}$ at time $t=1$ be drawn from a uniform distribution, e.g. from 5,000 to 50,000. Then for each bank set a conservative equity-to-assets ratio selected at random from the interval 0.10 to 0.35. And finally, let liquidity $L^{i}_{1}$ at $t=0$ be a fixed share of assets, thus featuring that a given bank can be solvent to provide loans or take additional debt, but still illiquid to do so. These steps so far are sufficient to characterize the banking system at the initial period.

\item[3.] Next assume the corporate sector requests loans from the banking system according to a Poisson process of rate $\lambda$, i.e. the length of time between loan requests is exponentially distributed. Let $\ell$ be the fixed amount of the loan, and $r_{C}$ be the interest rate of these loans. Each loan is supplied by only one bank $i$, where $i$ is chosen at random with equal probability among the $N$ banks. However, if bank $i$ is selected and $L^{i}_{{t-1}}<\ell$, it lacks enough liquid assets to fulfill the entire loan. To cover the difference, bank $i$ enters into the interbank market and borrows at rate \emph{r$_{B}$} (with $r_{C}>r_{B}$) from bank $j$, where $j$ is also selected randomly across the $N-1$ remaining banks.

\item[4.] Similarly, if $L^{j}_{t-1}<\ell-L^{i}_{t-1}$ then bank $j$ cannot supply the interbank loan. If this occurs then a new bank is chosen from the remaining $N-2$ banks.

\item[5.] As soon as $\ell$ is supplied to the corporate sector, liquidity $L^{m}_{t}$ as well as household deposits $D^{m}_{t}$ increase in a subset of banks $m \in M$ by the amount $\omega^{m}_{t}\times \ell$, where $\omega^{m}$ are random weights such that $\sum_{m}\omega^{m}_{t}=1$, and $M<N$. The intuition is that the corporate sector distributes the loan in the economy, for example in the form of wages, and households in turn deposit these proceeds randomly across a small number of banks. Notice that this generates balance sheet (assets) growth, which will further enrich the leverage dynamics. Although the $M$ banks are drawn randomly each time a loan takes place, $i$'s weights are fixed but re-scaled, i.e.  $\omega^{m}_{t}=\omega^{m}/\sum_{m^{\prime}}\omega^{m^{\prime}}$ $\forall{t},\forall{m,m'}\in M \subset N$. The rationale is that some banks tend to have a higher (or lower) share of deposits that is stable over time.

\item[6.] In addition, we propose that in each period there is a (low) probability of a negative shock to a random bank $i$. When the shock event is triggered, deposits and liquidity decrease in fixed amount, e.g. $0.2\times \ell$.\footnote{This type of liquidity-related credit shock is somewhat different from those in \cite{gai} or \cite{nier} since we are more interested in simulating lending dynamics rather than solvency or default risk per se.}

\item[7.] Finally, the corporate and interbank loans are paid back after $T$ periods. Therefore at time $t+T$ bank equity $E^{i}_{t+T}$ will increase by the net amount of interest earned between the loan granted to the corporate sector and, if any, the funds borrowed from the interbank market. Contrary to the loan being initially distributed across the banking system, the funds for repayment come exogenously from the real sector. In other words, we implicitly assume the corporate sector is always productive and therefore is able to generate proceeds elsewhere.
\end{enumerate}

Notice that the complete set of bank interactions resulting from the model can be described in the form of a directed network. In this graph nodes are given by the $N$ banks, and the directed links between nodes are given by the direction of the interbank loan. Therefore each point in time can be fully characterized by an adjacency matrix $\mathcal{A}^{t}$, where $\mathcal{A}^{t}$ is a $N\times N$ matrix and in general will not be symmetric. An entry $\mathcal{A}_{i,j}=1$ (i.e. edge from $i$ to $j$) implies bank $j$ has borrowed from bank $i$. However, we will not use this information for analyzing the data, we will only apply the methodology described before for the simulated Leverage time series to study the generated functional networks.

\subsection{Simulation Results}
\begin{figure}
\hspace*{-0.6cm}
\includegraphics[angle=0,width=1.05\textwidth]{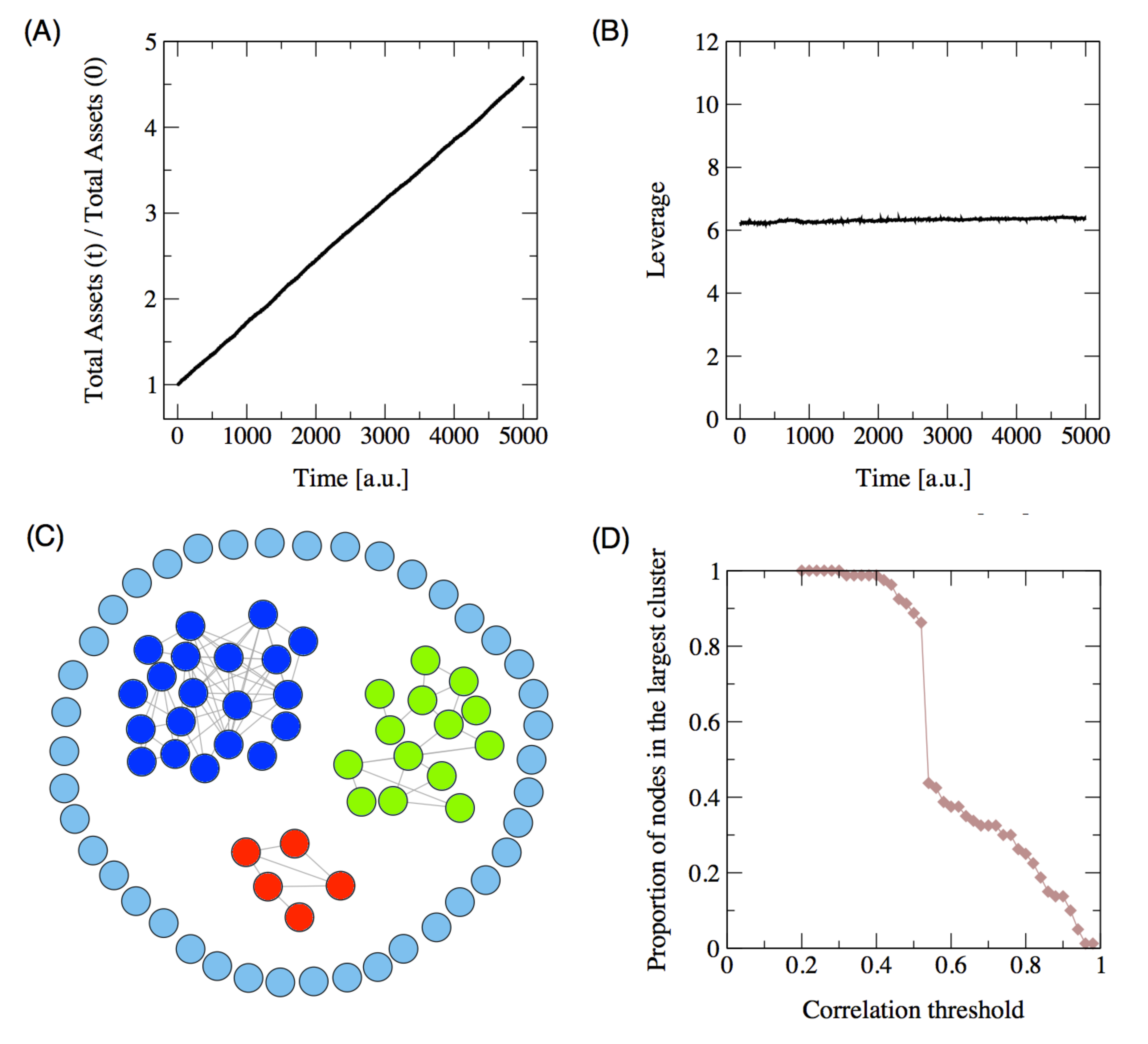} 
\caption{ \textit{Results for the simulated model with $N=80$ and 5,000 periods.}
(A) Banks' assets growth. (B) Average leverage over time. (C) Representation of simulated network. (D) Proportion of banks that belongs to the largest cluster as a function of the correlation threshold.}
\label{fig:simulacion1}
\end{figure}

We now simulate the model setting $N=80$ to be comparable to the actual size of the Argentine and Brazil network. This allows, without loss of generality, to more easily contrast the results with those in Section 3. In fact, results remain qualitatively similar if we modify parameters such as $N$, $\ell$, $r_C$, $r_B$, etc. For instance, consider Panel (A) and (B) of Fig. \ref{fig:simulacion1} which show the simulated dynamics of average assets and leverage over time, respectively. After some transient period (not shown) the average leverage become stationary (Panel (B)). Moreover, the stationary value is 6 which is very similar to the value observed in Argentina and Brazil (see Fig. 2). The average assets grows over time (Panel (A)), obtaining a growth of 4.5 for the time period studied. This last value is compatible to the Brazil assets growth (Fig. A1).  Now, that the average assets and leverage are similar to the empirical one, we study the individual bank behavior applying the network methodology described before for the individual simulated leverage time series.

Panel (C) in Fig. 7 shows the graphical representation of the simulated leverage network for a correlation threshold $\rho=0.8$. The topology is very similar to the network in Fig. 4. In particular, the network is characterized by a large interconnected cluster, together with smaller communities of banks, and many isolated banks. Furthermore, not only banks tend to merge with the largest cluster as $\rho$ decreases, but also it presents the discontinuity property that leads to a unique large cluster (Fig. 7 (D)). The proportion of nodes that belongs to the largest cluster jumps from 0.4 to over 0.8 at around $\rho=0.5$. The result shown in Panel (D) is very similar to that obtained in Fig. 5.

Therefore the interbank model generates modular structures that resemble the empirical networks. But can we say more about the mechanisms that drive the links between banks? In particular, we want to know whether the model can shed light on the dynamics behind the leverage dependence. In order to answer this we propose to study two banks, and in particular, the two most correlated banks. To avoid drawing conclusions from one simulation, we instead simulate the model 40 times, and for each of them we identify the two banks with the largest leverage dependence. Then compute the total leverage and assets growth, that is, the growth between the last period $t=5,000$ and $t=0$. We compute this ratio for both banks in each of the 40 replications.

The histogram of these ratios is depicted in panels (A) and (B) of Fig. 8. Interestingly, we find that the two banks exhibit large growth in both variables. In the model this takes place when banks receive greater frequency of loan requests. And at the same time, since these banks tend to have sufficient liquidity, they are also more likely to lend to other banks. The model therefore suggests that banks with more aggressive growth strategies may be more interconnected, and hence share additional risks between them.

We then translate this measure to the analog of the empirical samples. In Panels (C) and (D) of Fig. 8 we compute the leverage and the total assets growth for each bank in Argentina and Brazil, respectively. Where banks are ordered for better visualization. Recall we are comparing the simulations against Argentina and Brazil because they share a similar size $N=80$. Red and green dots identify the two most correlated banks in each country. Notice that both banks indeed exhibit a higher leverage growth. A higher assets growth is also found in Argentina, and to a less extent in Brazil's network.

\begin{figure}
\hspace*{-0.6cm}
\includegraphics[angle=0,width=1.05\textwidth]{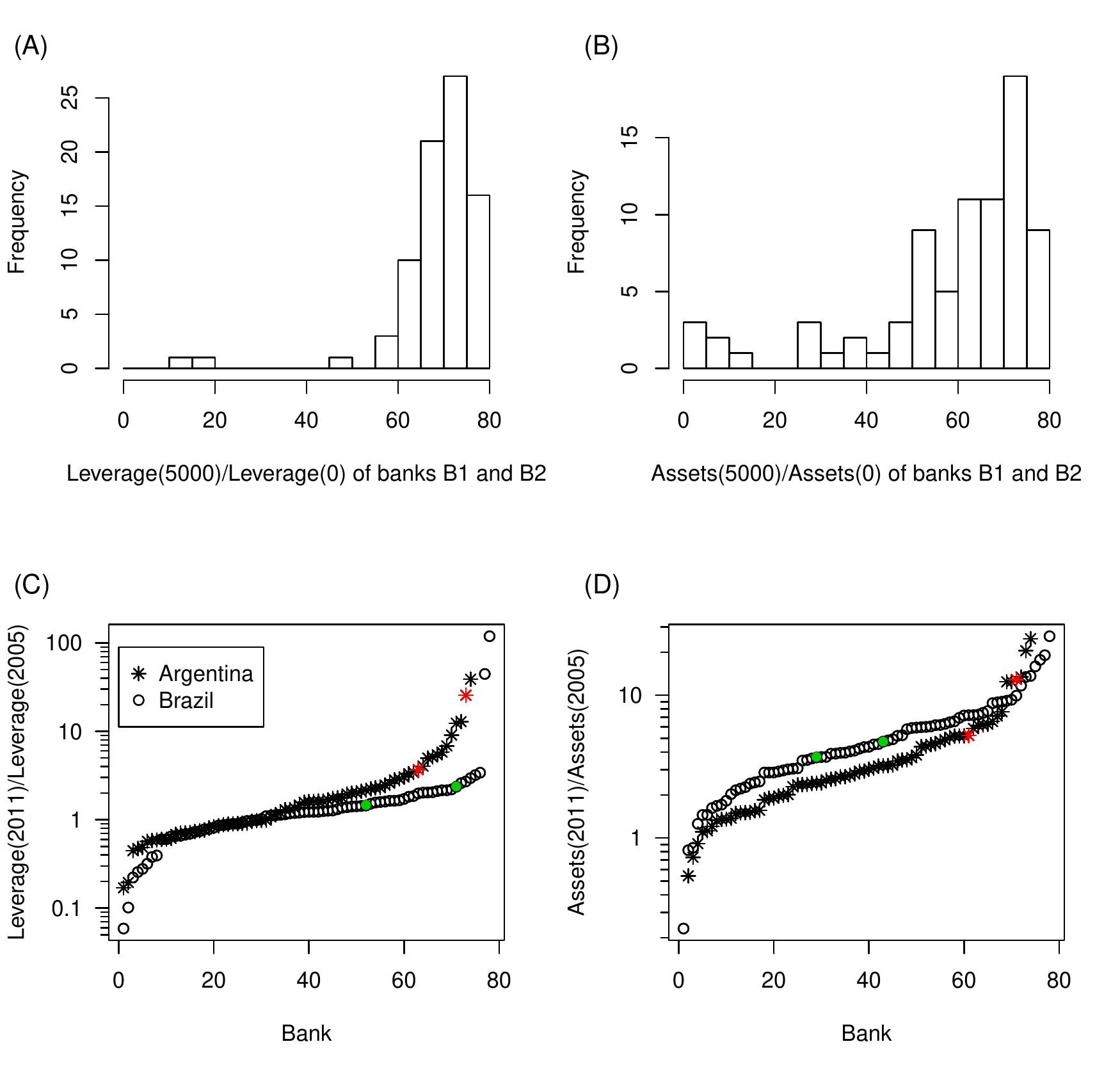} 
\caption{ \textit{Leverage and assets growth for the simulated and actual networks.}
(A) and (B) Leverage and assets growth for the  two most correlated banks, B1 and B2. (C) and (D) Leverage and assets growth in Argentina' and Brazil's networks.}
\label{fig:simulacion2}
\end{figure}

\section{Conclusion}

After the recent and still ongoing financial crisis an important bulk of research has turned its eyes into new ways to model financial risk and contagion between financial institutions, and how to improve regulation so as to reduce risk at both the individual and aggregate level. At the time of writing, banking supervision codes are being modified in order to incorporate tighter capital and liquidity requirements such that financial institutions are much better prepared to absorb losses in future credit-related events.\footnote{For an overview of the banking supervision regulation, see for example \cite{crisis1,crisis2,crisis3}.} This is particularly important since a crisis starting at a given place can very easily amplify or spread across the system through various channels: counterparty risk due to interbank lending, generalized fall in asset prices (fire sales to boost capital/margin requirements), liquidity hoarding or funding constraints (higher margins, run on repo/money market), self-fulfilling prophecies (bank runs), non-government bailout, or underestimation of risk in, and exposure to, derivate securities.

This paper studies the banking system within five different countries, and then compares the results across them.  We construct networks using bank-level balance sheet data, where links between each pair of banks are assigned upon the leverage ratio dependence. Although this interaction does not imply an interbank financial transaction, our findings do illustrate that recurring leverage dynamics (e.g. cluster of banks that simultaneously leverage or deleverage) is relevant in the understanding of risk concentration.

Interestingly, we find that in all countries (Argentina, Brazil, Mexico, South Africa, and Taiwan) the network topology is characterized by a large cluster of interconnected banks, together with other small communities and isolated banks. As the correlation threshold gets lower, the proportion of banks that belongs to the largest cluster increases almost dramatically: in both Argentina and Brazil, for instance, 90\% of the banks group into a unique cluster at a 0.5 threshold. We also find that these networks are not dominated by a hub (extremely interconnected node), but by a modular structure where communities are connected by bridge banks. This novel finding can be thought of as an optimal strategy to diversify risk within a group of banks that share similar characteristics. In other words, this subset restricts the spread of a distress cascade when one bank fails, as well as limits ex-post government assistance to a reduced set of banks.

To test the robustness of our results we present a model of corporate and interbank loans. The model generates rich balance sheet and lending dynamics, which certainly do not intend to capture all the intricacies between the financial system and the rest of the economy. Our objective is much smaller, and is to generate lending dynamics that can shed light on the complexity of interbank relationships. Interestingly, in doing so we obtain leverage networks that closely follow the empirical ones. Our simulations also suggest that banks with recent rapid or aggressive growth strategies are more likely to be interconnected, and hence share more risks.

Understanding risk and leverage using a network framework can also be of interest to policymakers and regulators, who for instance can build upon these tools to locate heavily interconnected banks and evaluate its importance to the aggregate system. When credit events occur, the indebtedness ratio is often the most straightforward and commonly used proxy to assess sustainability.
Policymakers should enrich this analysis by incorporating additional indicators of direct exposure such as interbank loans or repo agreements, which can also be used to construct more detailed networks.

Finally, understanding how individual or common shocks trigger financial crises is not only necessary to design countercyclical or containment policies, but also to design policies that prevent or limit the surge of such crises in the first place. As suggested in \cite{gai}, \cite{arinaminpathy} and \cite{upper}, institutions that are more likely to destabilize the system could be subject to stricter supervision, higher capital requirements or lower interbank exposure. This situation would certainly reduce moral hazard incentives for excessive risk-taking at institutions that are too large to be risked to fail. The G-20, the Financial Stability Oversight Committee and several countries on their own are already in the process of drafting special supervision guidelines for these so-called ``systematically important financial institutions" or SIFIs (\cite{bair} and \cite{elliot}).

\section*{Acknowledgements}
We are grateful to staff members from the corresponding Central or Reserve Banks for providing us the data.
We thank D. Heymann, J.C. Hallak, and two anonymous referees for helpful comments. 
\vspace*{-3pt}   



\begin{thebibliography}{32}
\bibitem{blanchard} Blanchard, O. J. ``The crisis: basic mechanisms and appropriate policies'', \emph{IMF Working Papers}, April, 1- 22 (2009).
\bibitem{duffie} Duffie, Darrell.Innovations in Credit Risk Transfer: Implications for Financial Stability.\emph{BIS Working Paper} 255 (2008).
\bibitem{haldane} A.G. Haldane, R.M. May. Systemic risk in banking ecosystems. \emph{Nature} 469:351-355 (2011).

\bibitem{buiter} Buiter, Willem, ``Lessons from the 2007 Financial Crisis'', Background Paper Submitted to the UK Treasury Select Committee, December 11 (2007).

\bibitem{gorton}Gorton, G.B. and Metrick, A. . ``Securitized banking and the run on repo''. Yale ICF Working Paper No. 09-14 (2010).

\bibitem{brunner} Brunnermeier, M.K.  ``Deciphering the liquidity and credit crunch 2007-2008''. \emph{Journal of Economic Perspectives}, 23, 1, 77-100 (2009).

\bibitem{klodt} Klodt H, Lehment H, eds. ``The Crisis and beyond. Lessons from the 2007 Financial Crisis''. Kiel Institute for the World Economy, E-Books, (2009).

\bibitem{fender}Fender, Ingo, and Jacob Gyntelberg. ``Overview: Global Financial Crisis Spurs Unprecedented Policy Actions''. \emph{BIS Quarterly  Review}, December (2008).

\bibitem{batti}     Battiston, S., Gatti, D.D, Gallegati, M., Greenwald, B.C.N., and Stiglitz, J.E. ``Liaisons dangereuses: increasing connectivity, risk sharing and systemic risk''. \emph{Journal of Economic Dynamics and Control}, 36, 1121-1141 (2012).

\bibitem{shin}  Shin, Hyun Song. ``Reflections on Northern Rock: The Bank Run That Heralded the Global Financial Crisis''. \emph{Journal of Economic Perspectives}, 23(1): 101-19 (2009).

\bibitem{basel1}   Basel Committee on Banking Supervision. Basel II: International Convergence of Capital Measurement and Capital Standards: a Revised Framework (2004).
\bibitem{basel2} Basel Committee on Banking Supervision. Basel III: A global regulatory framework for more resilient banks and banking systems (2010).

\bibitem{basel3} Hannoun, H. The basel III capital framework: a decisive breakthrough. Bank for International Settlements, (2010).

\bibitem{heise}
S. Heise and R. Kuhn . ``Derivatives and credit contagion in interconnected networks''. \emph{The European Physical Journal B - Condensed Matter and Complex Systems}, 85, Number 4, 115 (2012).
\bibitem{bastos}
Bastos e Santos, E. and Cont, R.. ``The Brazilian interbank network structure and systemic risk''. \emph{Banco Central do Brasil}, Working Paper No. 219 (2010).
\bibitem{nier} Nier, E., Yang, J., Yorulmazer, T., and Alentorn, A. ``Network Models and Financial Stability''. \emph{Journal of Economic Dynamics and Control}, 31(6): 2033-2060 (2007).
\bibitem{iori} Iori, G., Jafarey, S., and Padilla, F. ``Systemic risk on the interbank market''. \emph{Journal of Economic Behaviour and Organisation}, 61(4): 525-542 (2006).
\bibitem{kiyotaki} Kiyotaki, N. and Moore, J.. ``Credit cycles''. \emph{Journal of Political Economy}, 105, 2, 211-248 (1997).
\bibitem{krugman} Krugman, P. ``Balance sheets, the transfer problem, and financial crisis''. \emph{International Tax and Public Finance}, 6, 459-472 (1999).
\bibitem{cespedes2004} L.F. C\'espedes, R. Chang, A. Velasco. ``Balance sheets and exchange rate policy''. \emph{American Economic Review} {\bf 94}, 1183-1193 (2004).
\bibitem{mishkin1995} F.S. Mishkin. ``Symposium on the monetary transmission mechanism''. \emph{The Journal of Economic Perspectives} {\bf 9}, 4, 3-10 (1995).

\bibitem{kishan2000} R.P. Kishan, T.P. Opiela. ``Bank size, bank capital, and the bank lending channel''. \emph{Journal of Money, Credit, and Banking} {\bf 32}, 1, 121-141 (2000).
\bibitem{morris2008} S. Morris, H.S. Shin.``Financial regulation in a system context''. \emph{Brookings Papers on Economic Activity}, 229-274 (2008).
\bibitem{oliveira2011} R. Oliveira, R.F. Schiozer, L.A.B. Barros. ``Too big to fail perception by depositors: an empirical investigation''. \emph{Banco Central do Brasil}, Working Paper No. 233 (2011).
\bibitem{schmukler2001} M.S. Martinez Peria, S.L. Schmukler. ``Do depositors punish banks for bad behavior? Market discipline, deposit insurance, and banking crises''. \emph{The Journal of Finance} {\bf 56}, 3, 1029-1051 (2001).
\bibitem{maechler2006} A.M. Maechler, K.M. McDill. ``Dynamic depositor discipline in US banks''. \emph{Journal of Banking and Finance} {\bf 30} 7, 1871-1898 (2006).
\bibitem{imai2006} M. Imai. ``Market discipline and deposit insurance reform in Japan''. \emph{Journal of Banking and Finance},  30, 3433-3452 (2006).
\bibitem{flannery} M.J. Flannery. ``Using market information in prudential bank supervision: a review of the U.S. empirical evidence''. \emph{Journal of Money, Credit and Banking} {\bf 30}, 3, 273-305 (1998).
\bibitem{yeyati2004} E. Levy-Yeyati, M.S. Martinez Peria, S.L. Schmukler. ``Market discipline under systemic risk: evidence from bank runs in emerging economies''. \emph{The World Bank}, Policy Research Working Paper Series 3400 (2004).
\bibitem{galiani2003} S. Galiani, D. Heymann, M. Tommasi.``Great expectations and hard
times: the Argentine convertibility plan''. \emph{Econom\'ia} {\bf 3}, 2, 109-147 (2003).
\bibitem{calvo2003} G.A. Calvo, A. Izquierdo, E. Talvi. ``Sudden stops, the real Exchange
rate, and fiscal sustainability: Argentina's lessons''. NBER Working Paper No. 9828 (2003).
\bibitem{amato2002} L. D'Amato, T. Burdisso, V. Cohen. ``The Argentine banking and exchange rate crisis of 2001: can we learn something from financial crises?''. \emph{Banco Central de la Rep\'ublica Argentina}, Working Paper (2002).

\bibitem{newman} M. Newman, A.L. Barabasi, and D.J. Watts. ``The Structure and Dynamics of Networks''. Princeton Studies in Complexity, (2006).
\bibitem{caldarelli} G. Caldarelli. ``Scale-Free Networks: Complex Webs in Nature and Technology''. Oxford Finance, (2007).
\bibitem{crisis1} Bank for International Settlements. Global systemically important banks: assessment
methodology and the additional loss absorbency requirement. (2011)
\bibitem{crisis2}Bank of England, The role of macroprudential policy. Discussion paper, November (2009).
\bibitem{crisis3}Bank of England, Instruments of macroprudential policy. Discussion paper, December (2011).

\bibitem{raberto} M. Raberto, F. Rapallo, and E. Scalas. ``Semi-Markov graph dynamics''. PloS one 6.8: e23370, (2011).
\bibitem{chen1999} Y. Chen. ``Banking panics: the role of the first-come, first-served rule and information externalities''. \emph{Journal of Political Economy}, 107 (5), 946–968, (1999).
\bibitem{lengnick} M. Lengnick, S. Krug, and H.W. Wohltmann. ``Money Creation and Financial Instability: An Agent-Based Credit Network Approach''. Economics: The Open-Access, Open-Assessment E-Journal, 7 (2013-32): 1-44, (2013).
\bibitem{lebaron} B. LeBaron, and L. Tesfatsion. ``Modeling Macroeconomics as Open-Ended Dynamic Systems of Interaction Agents''. \emph{American Economic Review}, 98(2): 246-50, (2006).


\bibitem{gai} Gai, P., Haldane, A., Kapadia, S. ``Complexity, concentration and contagion''. \emph{Journal of Monetary Economics}, 58(5), 453-470 (2011).
\bibitem{arinaminpathy} Arinaminpathy, N., Kapadia, S.,  May, R. M.. ``Size and complexity in model financial systems''. \emph{Proceedings of the National Academy of Sciences}, (2012).
\bibitem{upper} Upper, C, ``Simulation methods to assess the danger of contagion in interbank markets''. \emph{Journal of Financial Stability}, 7(3), 111-125 (2011).
\bibitem{bair}  Bair, S.C. ``We must resolve to end too big to fail''. Proceedings, Federal Reserve Bank of Chicago, May, 9-15 (2011).
\bibitem{elliot}   Elliot, D. J. and Litan, R. E. ``Identifying and Regulating Systemically Important Financial Institutions: The Risks of Under and Over Identification and Regulation''. Washington: The Brookings Institution, (2011).
%












\end{thebibliography}

\begin{center}
\line(1,0){250}
\end{center}
\newpage

\section*{Supplementary data}

\setcounter{figure}{0}
\renewcommand{\figurename}{Figure A}



\begin{figure}[h]
\hspace*{-0.5cm}
\includegraphics[width=0.9\textwidth]{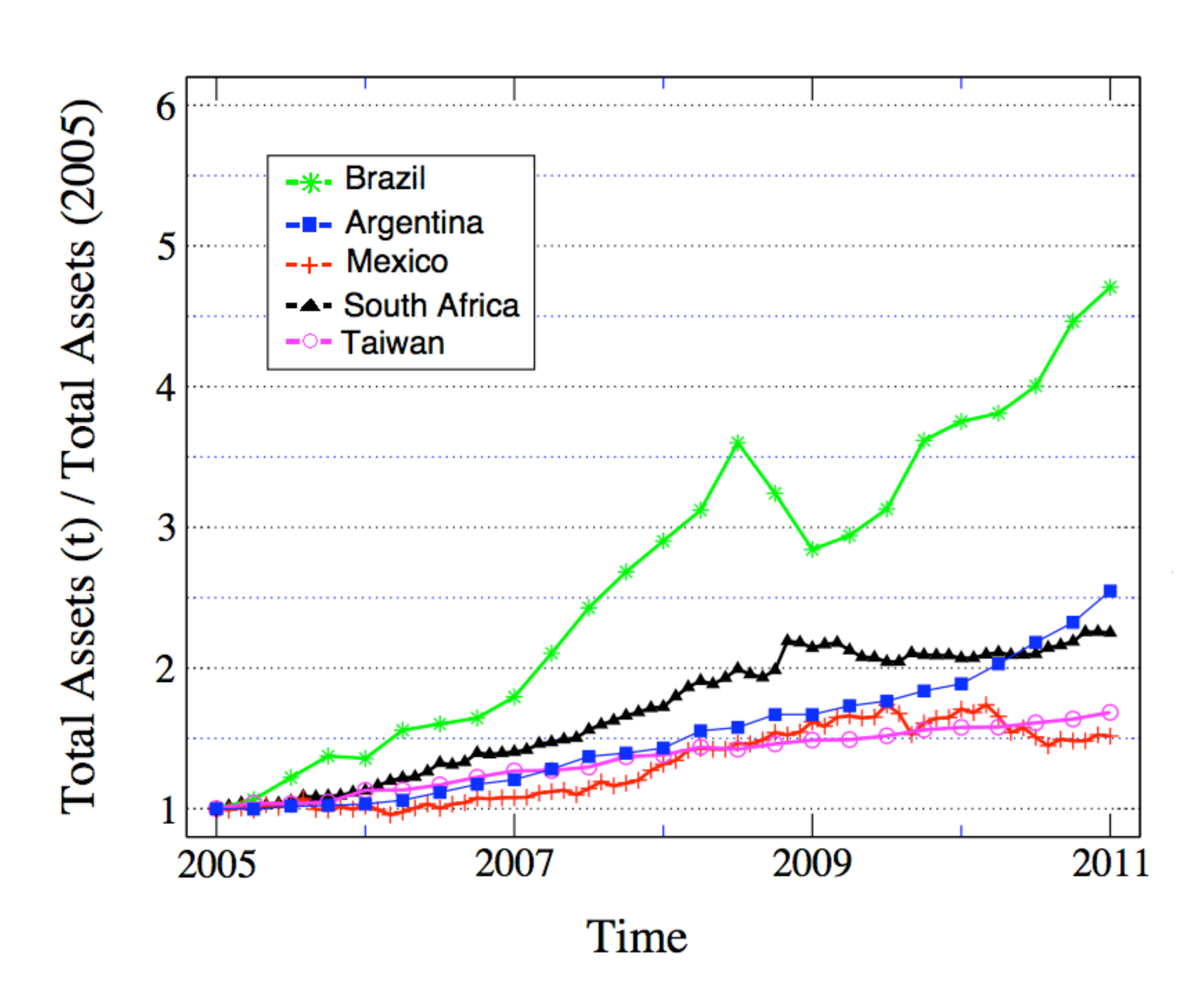} 
\caption{\textit{Banks' assets over time.} Total assets as a function of time for Argentina,  Brazil, Mexico, South Africa, and Taiwan.  For each country total assets was normalized by its value in 2005.}
\label{fig:activos}
\end{figure}

\end{document}